\documentclass[%
preprint, 
 amsmath,amssymb,
 aps, physrev,
]{revtex4-2}

\usepackage{graphicx}
\usepackage{dcolumn}
\usepackage{bm}

\begin{document}

\preprint{APS/123-QED}

\title{\textbf{Quantum Limits to Ground-State Cooling of Traveling Hypersound Phonons} 
}%

\author{Juntong Yang}
\email{Contact author: jyang298@uottawa.ca}
\affiliation{%
 Department of Physics, University of Ottawa, Ottawa, Ontario K1N 6N5, Canada
}%
\affiliation{%
 Nexus for Quantum Technologies Institute, University of Ottawa, Ottawa, Ontario, Canada
}%

\author{Liang Chen}
\affiliation{%
 Department of Physics, University of Ottawa, Ottawa, Ontario K1N 6N5, Canada
}%
\affiliation{%
 Nexus for Quantum Technologies Institute, University of Ottawa, Ottawa, Ontario, Canada
}%
 
\author{Xiaoyi Bao}%
 \email{Contact author: Xiaoyi.Bao@uottawa.ca}
\affiliation{%
 Department of Physics, University of Ottawa, Ottawa, Ontario K1N 6N5, Canada
}%
\affiliation{%
 Nexus for Quantum Technologies Institute, University of Ottawa, Ottawa, Ontario, Canada
}%



\begin{abstract}
The steady final phonon occupation in waveguide optomechanical systems based on backward stimulated Brillouin-Mandelstam scattering has not been established in the strong-coupling regime. In this work, the displacement spectra of anti-Stokes optical modes and acoustic modes in tapered chalcogenide photonic crystal fiber are derived from the Lindblad (or Gorini-Kossakowski-Sudarshan-Lindblad) master equation. By analyzing the full spectral response, we indicate that the system can enter the strong-coupling regime through the emergence of normal-mode splitting and avoided crossings. Within a non-Hermitian framework, the threshold for strong coupling is identified, showing that it can be achieved at relatively low pump power even at room temperature. Furthermore, we derive a unified analytical expression for the final phonon occupation, revealing that quantum backaction and zero-point fluctuations impose additional fundamental limits that hinder the achievement of ground-state cooling. These results redefine the quantum limits of steady-state cooling in continuum optomechanics, motivating the search for new strategies to access the quantum ground-state of macroscopic phonons.

\end{abstract}

\maketitle


\section{\label{sec:level1}Introduction}
Cooling of thermal motion plays a critical role in enhancing the performance of precision measurement systems, including gravitational-wave detectors \cite{whittle2021approaching, gras2017audio, abbott2009observation, gonzalez2000suspensions}, optical atomic clocks \cite{Reilly2026fully, liu2018orbit, ludlow2015optical, kessler2012sub, bize2005cold}, quantum computing platforms \cite{odeh2025non, bassman2024dynamic, fabrikant2024cooling, feng2022quant}, and ultrasensitive nanobolometers for sub-millimeter telescopes \cite{koppinen2009phonon, wei2008ultrasens}. On the other hand, laser-based cooling has emerged as a powerful approach for reaching the quantum ground state, enabling applications such as quantum state preparation \cite{clements2026sub, teufel2011sideband, riviere2011optom, jaehne2008ground} and fundamental research including cavity quantum electrodynamics \cite{Birnbaum2025photon}, quantum optomechanics \cite{diamandi2025optom, marshall2003towards}, Casimir effect \cite{ferreri2024phonon}. 

The theoretical framework and experimental realization of radiation-pressure cooling were initially developed in resonant cavity systems \cite{aspelmeyer2014cavity, braginski1967ponder, dorsel1983optical}. Over the past two decades, advances in nanofabrication have powered rapid progress in quantum cavity optomechanics. In these systems, the interaction between light and mechanical motion gives rise to Stokes and anti-Stokes scattering processes, corresponding to heating and cooling, respectively. Efficient cooling toward the quantum ground state requires enhancement of the anti-Stokes process while suppressing the Stokes process. This condition is fulfilled when the cavity linewidth is sufficiently narrow to spectrally distinguish the two processes, which defines the resolved sideband regime \cite{aspelmeyer2014cavity, kippenberg2008cavity}. 

Within this regime, the optomechanical interaction can be further characterized by the relative strength of the coupling rate and the optical dissipation rates. In the weak coupling condition, the final phonon number is determined by the competition between thermal heating and optical cooling \cite{marquardt2007quantum, wilson2007theory}. Coupling to the thermal environment drives the mechanical mode toward its equilibrium occupation $n_{\mathrm{th}}$, while coupling to the optical field cools the mechanical motion but is fundamentally limited by quantum backaction. This backaction originates from residual fluctuations associated with Stokes scattering, which introduces heating and sets a lower bound $n_{\mathrm{min}}$ on the achievable phonon number. As such, the steady-state occupation accounts for the balance between these two dissipation channels. 

However, this description assumes that the optomechanical interaction can be treated as a perturbative dissipative process \cite{aspelmeyer2014cavity}. When the coupling strength becomes comparable to the optical dissipation rates, the system enters the strong coupling regime, where coherent energy exchange between photons and phonons dominates the dynamics. In this regime, the eigenmodes hybridize into mixed optical and mechanical excitations, and the picture of independent coupling to thermal and optical reservoirs breaks down. Therefore, the final phonon number can no longer be described by a simple balance between two dissipation channels, but is instead governed by the full spectral response of the coupled system, manifesting in phenomena such as normal-mode splitting (NMS) \cite{marquardt2007quantum, dobrindt2008parametric, wilson2008cavity}. 

In contrast to cavity optomechanical systems, which typically address discrete mechanical modes, waveguide optomechanics enables the cooling of a continuum of traveling-wave phonons with high frequencies. This cavity-less approach has been theoretically proposed \cite{chen2016brillouin, zhu2023dynamic} and demonstrated experimentally in silicon waveguides \cite{otterstrom2018om} and optical fibers \cite{johnson2023laser, blazquez2024oa}. The underlying mechanism relies on backward stimulated Brillouin-Mandelstam scattering (SBS), in which Stokes and anti-Stokes processes are mediated by distinct phase-matched phonon modes. As a result, this phase-matching condition intrinsically breaks the symmetry between heating and cooling. Unlike cavity optomechanics requiring sideband resolution, the momentum conservation in this mechanism ensures that the Stokes process does not directly repopulate the same phonon mode being cooled. Consequently, the minimum phonon occupation $n_{\mathrm{min}}$, which arises from quantum backaction in cavity systems, is absent in the waveguide configuration. In the weak coupling regime, the result of this analysis is consistent with the expression reported in Ref.~\cite{johnson2023laser, blazquez2024oa}, where the differences stem from the distinct forms of the optomechanical damping rate. Recent experiments have demonstrated strong optomechanical coupling in fiber-based continuous systems at cryogenic temperatures \cite{martinez2025cavity}. However, achieving this regime in waveguide optomechanical cooling platforms under room temperature remains an open challenge. In particular, it is not yet clear how the steady phonon occupation should be described once this regime is reached. 

Here, we theoretically derive the displacement spectra of anti-Stokes optical modes and acoustic modes by quantizing the backward SBS interaction under the undepleted pump condition. Based on these spectra, the cooling platform of tapered chalcogenide glass photonic crystal fiber (PCF) is capable of entering the strong coupling regime, as evidenced by the emergence of one key spectroscopic signature, NMS. A complementary non-Hermitian eigenmode analysis reveals an additional signature of strong coupling, avoided crossings, and allows us to identify the corresponding threshold, demonstrating that this regime is accessible at comparatively low pump power under room temperature. A unified analytical expression for the final phonon occupation is obtained via contour integration of the spectral response of the phase-matched acoustic mode. Our analysis indicates that quantum backaction and zero-point fluctuations set fundamental bounds that prevent reaching the ground state. These results provide a comprehensive theoretical framework of optoacoustic cooling of macroscopic phonons while establishing revised quantum limits for steady-state cooling in continuum optomechanics, challenging prior assumptions and pointing to the need for new pathways toward the ground-state of macroscopic phonons.

\section{Theoretical Model}
In the quantum mechanical description, the backward Brillouin anti-Stokes scattering can be interpreted as the creation of an anti-Stokes photon, accompanied by the simultaneous annihilation of a pump photon and an acoustic phonon. Under the undepleted pump approximation, this interaction can be treated as an open quantum subsystem coupled to external reservoirs. Starting from the total Hamiltonian of the subsystem, bath, and their interaction, and applying the Lindblad equation  (also called Gorini-Kossakowski-Sudarshan-Lindblad equation) in the Heisenberg picture, the linearized Heisenberg–Langevin equations in momentum space can be written as
\begin{subequations}
    \label{eq1}
    \begin{equation}
        \frac{d a}{d t} = \left(-\frac{\gamma_o}{2} + i\Delta_{as}\right)a -ig_{om} b + \sqrt{\gamma_o}\xi_{a}, \label{eq1a}
    \end{equation}
    \begin{equation}
        \frac{d b}{d t} = \left(-\frac{\Gamma_m}{2} + i\Delta_{ac}\right)b -ig_{om} a + \sqrt{\Gamma_m}\xi_{b}, \label{eq1b}
    \end{equation}
\end{subequations}
where $a(k,t)$ and $b(q,t)$ denote the bosonic annihilation operators for the $k$-th anti-Stoke photon mode and acoustic phonon mode, respectively. $\gamma_o$ and $\Gamma_m$ represent the optical and acoustic dissipation rates. $\Delta_{as} = kv_{o}$ and $\Delta_{ac} = qv_{ac}$ describe the wavevector-induced frequency detunings, where $v_{o}$ and $v_{ac}$ are the group velocities of the anti-Stokes and acoustic mode. Mode matching in the interaction requires $k = q$, with perfect phase matching achieved at $k=q=0$. $g_{om}$ is the pump-enhanced coupling strength. $\xi_a$ and $\xi_b$ are quantum Langevin noises associated with reservoirs and weighted with the rates $\gamma_o$ and $\Gamma_m$. Detailed derivations are provided in Appendix \ref{appA}. Since the inherent symmetry breaking between the anti-Stokes and Stokes processes in the backward SBS in waveguides, phonon cooling can be individually performed \cite{blazquez2024oa}. Hence, only annihilation operators appear in coupling terms of Eq.(\ref{eq1}), compared to the full interaction encountered in typical cavity optomechanics \cite{marquardt2007quantum, dobrindt2008parametric}. 

By moving into the frequency domain via Fourier transformation, the displacement spectrum of the $k$-th anti-Stokes mode is obtained as
\begin{equation}
    \begin{aligned}
        S_{XX}^a(k, \omega) = \frac{\frac{\gamma_o}{2} (2n_{\text{as}} + 1) \left[\frac{\Gamma_m^2}{4} + (\omega + \Delta_{ac})^2\right] + g_{om}^2 \frac{\Gamma_m}{2} (2n_{\text{th}}+1)}{\left| g_{om}^2 + \left[\frac{\gamma_o}{2} - i(\omega + \Delta_{as})\right]\left[\frac{\Gamma_m}{2} - i(\omega + \Delta_{ac})\right] \right|^2},
    \end{aligned} \label{eq2}
\end{equation}
where the thermal occupation of phonon $n_{\text{th}}$ with frequency $\Omega_{ac}$ at the environment temperature $T$ obeys the Bose-Einstein distribution, $n_{\text{th}} = \left[\text{exp}(\hbar \Omega_{ac}/k_BT) - 1\right]^{-1}$. $n_{\text{as}}$ denotes the effective occupation of the optical reservoir associated with the anti-Stokes field. It is defined through the noncommuting noise operators $\left\langle \xi_{as}^{\dagger}(t_1) \xi_{as}(t_2) \right\rangle = n_{\text{as}}\delta(t_1 - t_2)$, which encodes quantum vacuum fluctuations responsible for optomechanical backaction. This assumption is aligned with the picture of amplified spontaneous anti-Stokes scattering, which can be viewed as introducing a fictitious photon per mode in the system \cite{smith1972optical, Agrawal2013}. Details are provided in Appendix \ref{appB}. 

Similarly, the displacement spectrum for the $q$-th acoustic mode is given by
\begin{equation}
    \begin{aligned}
        S_{XX}^b(q, \omega) = \frac{\frac{\Gamma_m}{2} (2n_{\text{th}}+1) \left[\frac{\gamma_o^2}{4} + (\omega + \Delta_{as})^2\right] + g_{om}^2 \frac{\gamma_o}{2} (2n_{\text{as}} + 1)}{\left| g_{om}^2 + \left[\frac{\gamma_o}{2} - i(\omega + \Delta_{as})\right]\left[\frac{\Gamma_m}{2} - i(\omega + \Delta_{ac})\right] \right|^2}.
    \end{aligned} \label{eq3}
\end{equation}
The displacement spectrums characterize the frequency distribution of the optical and acoustic modes. As autocorrelation functions of the corresponding operators, their integrals determine the effective photon and phonon populations. We then introduce the optical and acoustic response functions $\chi_{as}(\omega) = \left[\frac{\gamma_o}{2} - i(\omega + \Delta_{as})\right]^{-1}$ and $\chi_{ac}(\omega) = \left[\frac{\Gamma_m}{2} - i(\omega + \Delta_{ac})\right]^{-1}$, which describe the linear susceptibility of the respective modes. 

The coupled equations Eq.(\ref{eq1}) can be recast in matrix form as
\begin{equation}
    \frac{d}{dt} \begin{pmatrix}
        a \\ b
    \end{pmatrix} = -i\begin{pmatrix}
        -\Delta_{as} - i\frac{\gamma_o}{2} & g_{om} \\
        g_{om} & -\Delta_{ac} - i\frac{\Gamma_m}{2}
    \end{pmatrix} \begin{pmatrix}
        a \\ b
    \end{pmatrix} + \begin{pmatrix}
        \sqrt{\gamma_o}\xi_{a} \\ \sqrt{\Gamma_m}\xi_{b}
    \end{pmatrix}. 
\end{equation}
The dynamics of this quantum subsystem is governed by a non-Hermitian Hamiltonian, capturing both coherent coupling and dissipative interactions with the environment \cite{miri2019exceptional, wang2023non, zhang2019quantum, el2018non}. By neglecting Langevin noise, the complex eigenvalues are extracted from the effective Hamiltonian as
\begin{equation}
    \omega_{\pm} = -\frac{\Delta_{as} + \Delta_{ac}}{2} -i\frac{\gamma_o + \Gamma_m}{4} \pm \sqrt{g_{om}^2 + \left(\frac{\Delta_{as} - \Delta_{ac}}{2} + i\frac{\gamma_o - \Gamma_m}{4}\right)^2}, \label{eq5}
\end{equation}
which define the poles of the displacement spectra. These poles appear in conjugate pairs $\omega_{\pm}$ and $\omega_{\pm}^\ast$, providing the basis for contour integration in the evaluation of final phonon populations. A key feature of non-Hermitian systems is the existence of exceptional points (EPs), where both eigenvalues and eigenvectors coalesce. In the present system, the EP marks the transition between weak and strong coupling regimes, beyond which the optical and acoustic modes hybridize and emerge spectral signatures,  NMS and avoided crossing. 

\section{Strong Coupling Regime}
In this section, we analyze the coupling strength of the system through the spectral features of the anti-Stokes optical field as a function of pump power. The waveguide cooling platform considered here is a tapered chalcogenide PCF, as demonstrated in Ref.\cite{blazquez2024oa}. At room temperature, the relevant experimental parameters are $\gamma_o = 364$ MHz, $\Gamma_m = 46.8$ MHz, the SBS gain coefficient $G_B = 164$ m$^{-1}$W$^{-1}$, and the core refractive index $n=2.5$. To ensure consistency with units of the dissipation rates, the optomechanical coupling strength is expressed as $g_{om} \approx \sqrt{G_B\Gamma_m P c/(4n)}$ \cite{martinez2025cavity, zhang2023quantum}, where $P$ is the peak power of pump pulse and $c$ is the light speed in vacuum. 

\subsection{The overall spectrum}
\begin{figure}[h!]
  \includegraphics[width = 13cm]{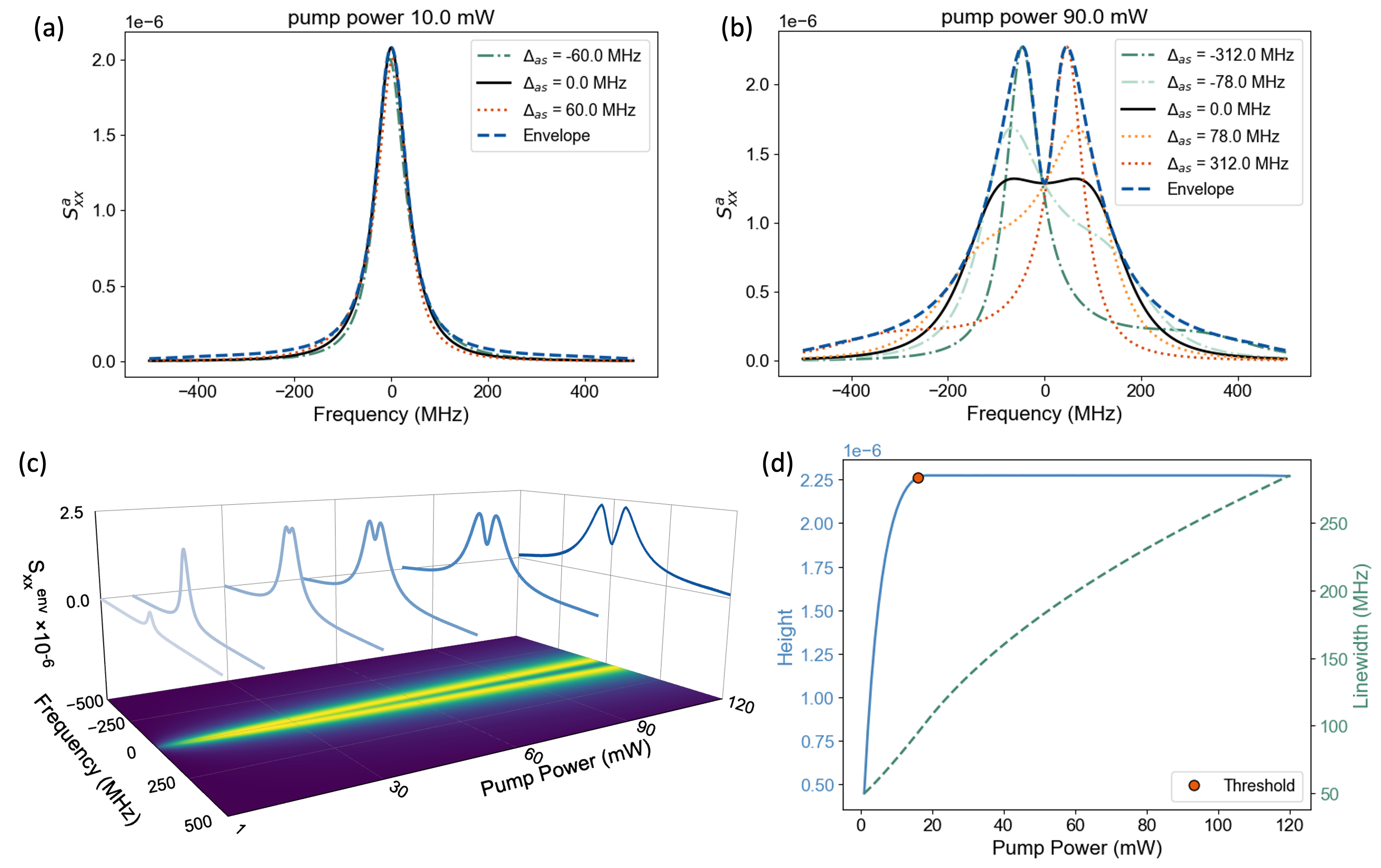}
  \caption{The overall spectrum envelope at (a) weak coupling regime and (b) strong coupling regime. (c) The spectrum envelope evolution and (d) the height and linewidth of envelope as the pump power. \label{figure1}}
\end{figure}

The overall spectral features of the anti-Stokes light field are analyzed using the displacement spectrum in Eq.(\ref{eq2}). At low pump peak power, the response is dominated by the perfectly phase-matched optical mode, as shown in Fig.\ref{figure1}(a). At this stage, the linewidth of the measured spectrum directly     characterizes the linewidth of the phase-matched mode. As the pump power increases, modes with finite phase mismatch begin to participate in the interaction, as illustrated in Fig.\ref{figure1}(b). For the phase-matched optical mode, the displacement spectrum remains symmetric with respect to frequency, reflecting the balanced response of the optical and acoustic modes. In comparison, finite phase mismatch introduces asymmetry in the spectrum due to the frequency-dependent interplay between the optical and acoustic susceptibilities. This asymmetry is nevertheless constrained by the underlying time-reversal symmetry, which ensures spectral mirror symmetry between $\pm \Delta_{as}$.

The total spectrum, formed by the superposition of all contributing modes, remains globally symmetric and develops a characteristic splitting into two peaks, know as NMS. This constitutes a clear spectroscopic signature of the strong coupling regime. In this regime, the concept of a single linewidth loses its physical meaning, as the spectrum is no longer governed by an isolated mode but instead reflects the hybridized response of the coupled system. As a result, it is more appropriate to evaluate the final phonon occupation from the full spectral response rather than from a single linewidth parameter.

Fig.\ref{figure1}(c) shows that with increasing pump power, both the separation between the two peaks and the depth of the central dip increase. The evolution of the overall spectral peak height and linewidth is summarized in Fig.\ref{figure1}(d). The peak height approaches saturation (reaching $99\%$ of its maximum value) at a pump power of $16$ mW, in agreement with the SBS threshold reported in Ref.\cite{blazquez2024oa}. The observed linewidth broadening results from two contributions: the intrinsic broadening of the phase-matched mode and the increasing participation of more off-resonant modes. By contrast, the saturation of the peak height is primarily governed by the latter effect.

\begin{figure}[h!]
  \includegraphics[width = 15.5cm]{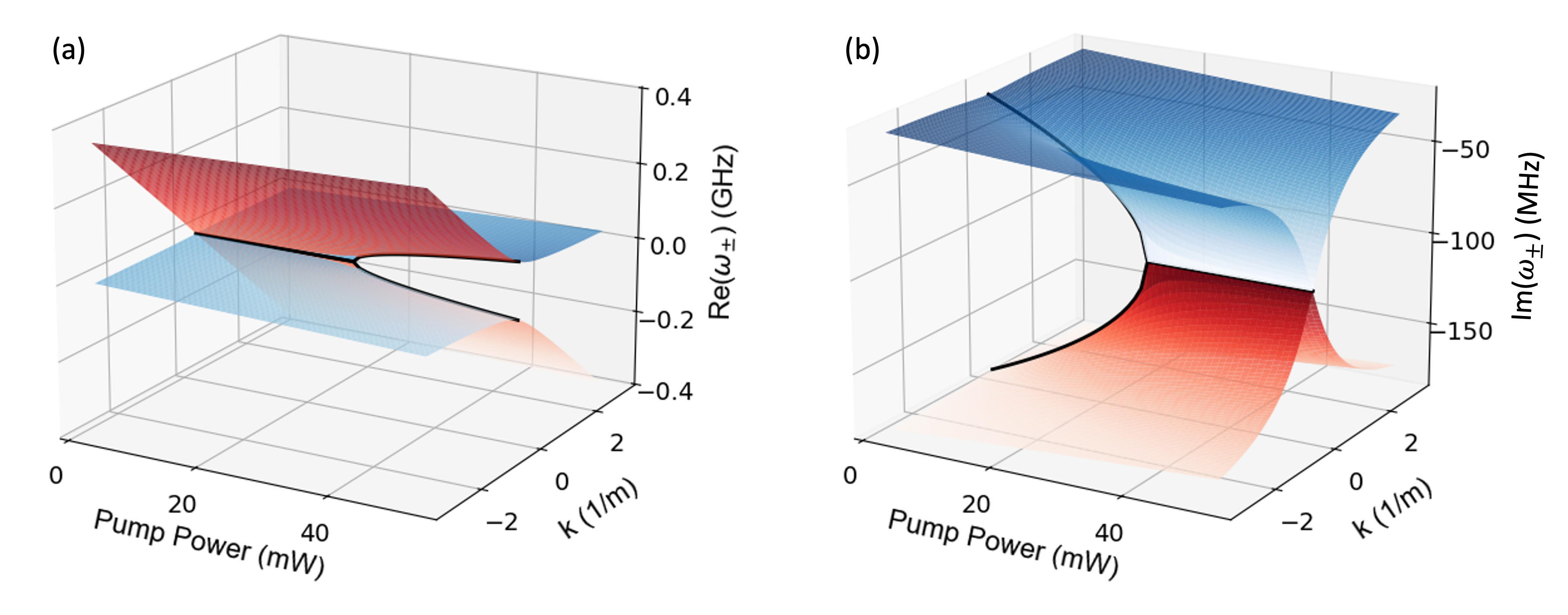}
  \caption{(a) The real part and (b) imaginary part of complex eigenvalues. Black lines correspond to the phase-matched modes. \label{figure2}}
\end{figure}

Another signature of the strong coupling regime is the appearance of avoided crossings in the system eigenfrequencies. This behavior can be analyzed using the Riemann surfaces defined by Eq.(\ref{eq5}) in a two dimensional parameter space. As shown in Fig.\ref{figure2}(a), when the pump power exceeds the EP, the eigenvalue branches exhibit an avoided crossing as a function of the phase-mismatch wavevector $k$. In addition, the spectral asymmetry associated with finite phase mismatch is evident in this representation. The system supports two distinct eigenmodes with different resonance frequencies (real parts) and linewidths (imaginary parts). At low pump power, one dominant eigenmode remains close to the central resonance frequency with only a small shift, while the other mode exhibits weak amplification. The phase-mismatched modes therefore appear slightly tilted relative to the phase-matched mode, as shown in Fig.\ref{figure1}(a). Upon entering the strong coupling regime, both the frequency shift of the dominant mode and the gain of the secondary mode increase significantly, showing their hybrid optical–acoustic character. 

\subsection{The perfect phase-matching mode}
\begin{figure}[h!]
  \includegraphics[width = 13cm]{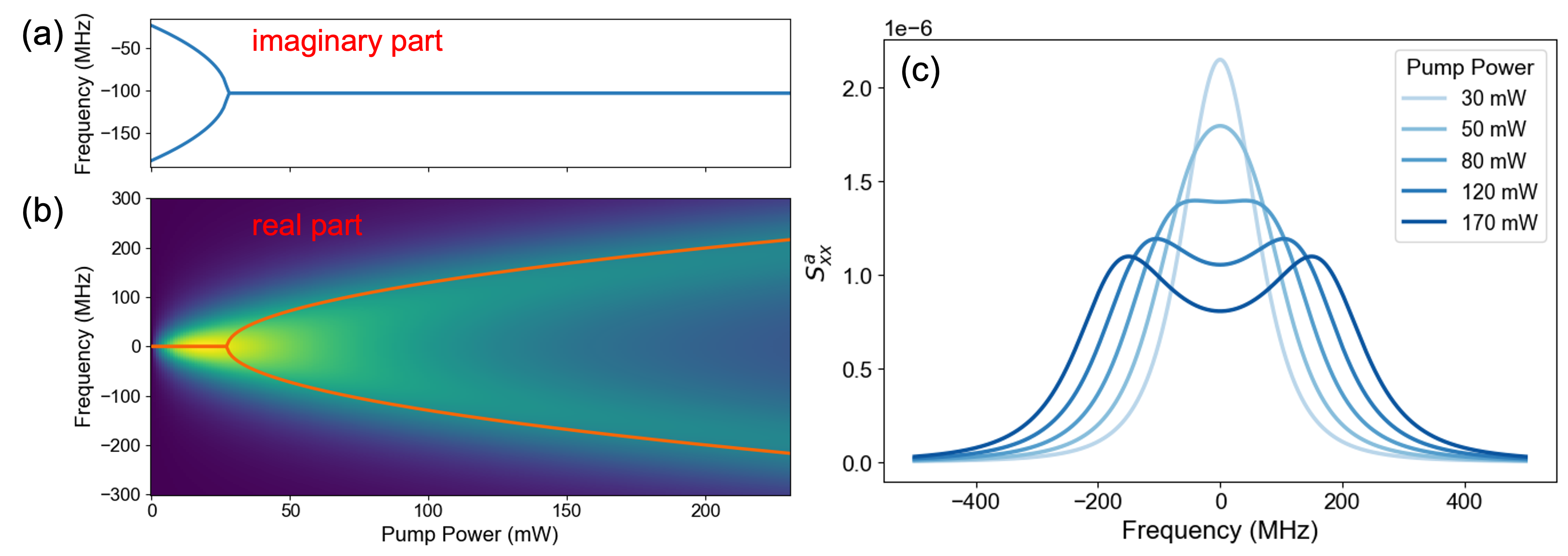}
  \caption{The perfect phase-matched mode. (a) The imaginary part of eigenvalues. (b) The real part of eigenvalues and the displacement spectrum $S_{XX}^a(k=0, \omega)$. (c) The displacement spectrum $S_{XX}^a(k=0, \omega)$ at different pump power. \label{figure3}}
\end{figure}

While the global spectral response provides clear phenomenological signatures of strong coupling, such as NMS and avoided crossings, a quantitative criterion for the onset of this regime can be obtained by examining the perfectly phase-matched optical mode. This mode dominates the interaction at low pump power and provides an explicit reference for identifying the EP and the corresponding strong coupling threshold, as suggested by Fig.\ref{figure2}. Under the perfect phase-matching condition, Eq.(\ref{eq5}) reduces to
\begin{equation}
    \omega_{\pm}^{pm} = -i\frac{\gamma_o + \Gamma_m}{4} \pm \sqrt{g_{om}^2 - \left(\frac{\gamma_o - \Gamma_m}{4}\right)^2},  \label{eq6}
\end{equation}
which directly reveals the transition between weak and strong coupling through the behavior of the square-root term. The displacement spectrum $S_{XX}^a(k=0,\omega)$ of the phase-matched anti-Stokes mode is shown in Fig.\ref{figure3}(b) and (c). The imaginary and real parts of the eigenvalues, obtained from Eq.(\ref{eq6}), are plotted in Fig.\ref{figure3}(a) and (b), which are consistent with the phase-matched branches highlighted in Fig.\ref{figure2}. 

In the stage $g_{om} < (\gamma_o - \Gamma_m)/4$, the system operates in the weak coupling regime. In this case, the eigenvalues remain purely imaginary, indicating that the dynamics are dominated by dissipation. The real parts of the eigenfrequencies are unchanged, while the effective decay rates are modified by the optomechanical interaction. As the coupling strength increases ($g_{om} \propto \sqrt{P}$), the amplitude of the anti-Stokes signal grows, describing enhanced phonon annihilation and improved cooling efficiency. 

When the coupling strength exceeds the critical value $g_{om} > (\gamma_o - \Gamma_m)/4$, the square root term in Eq.(\ref{eq6}) becomes real, signaling the transition to the strong coupling regime. In this regime, the eigenfrequencies split into two distinct branches, leading to NMS even for the phase-matched optical mode. Each mode exhibits a linewidth of $(\gamma_o + \Gamma_m)/4$, and the spectrum remains symmetric. As the pump power is further increased such that $g_{om} > \gamma_o$, the system enters the quantum coherent regime, where the rate of energy exchange between optical and acoustic modes exceeds the decoherence rate induced by the environmental quantum noise. In this regime, coherent photon–phonon dynamics dominate the system behavior \cite{dobrindt2008parametric, aspelmeyer2014cavity}.

Compared with the weak coupling regime, the peak amplitude of the anti-Stokes signal decreases with increases in pump power once entering strong coupling regime. This behavior indicates that cooling efficiency gradually decreases, marking the transition from dissipative cooling to coherent dynamics. The threshold pump power for entering the strong coupling regime is obtained from the EP condition as
\begin{equation}
    P_s = \frac{4n}{G_B\Gamma_m c}\left(\frac{\gamma_o - \Gamma_m}{4}\right)^2 \approx 27.3\ \text{mW},
\end{equation}
demonstrating that strong coupling can be achieved in waveguide optomechanical systems at room temperature with relatively low pump power.

\section{Limit of Cooling}
The above analysis shows that a tapered chalcogenide PCF platform can access the strong coupling regime at room temperature. Having established this, it is essential to revisit the description of cooling performance, as the phonon occupation need to be determined from the full spectral response of the coupled system in the strong coupling regime. For clarity, we restrict the analysis to the perfectly phase-matched acoustic mode, where the interaction is strongest and the analytical expressions are tractable. Under this condition, the displacement spectrum of the acoustic mode in Eq.(\ref{eq3}) can be rewritten as
\begin{equation}
    \begin{aligned}
        S_{XX}^b(q=0, \omega) &= \frac{\frac{\Gamma_m}{2} (2n_{\text{th}}+1) \left(\frac{\gamma_o^2}{4} + \omega^2\right) + g_{om}^2 \frac{\gamma_o}{2} (2n_{\text{as}}+1)}{\left| g_{om}^2 + \left(\frac{\gamma_o}{2} - i\omega\right)\left(\frac{\Gamma_m}{2} - i\omega\right) \right|^2} \\[5pt] &= \frac{\frac{\Gamma_m}{2} (2n_{\text{th}}+1) \left(\frac{\gamma_o^2}{4} + \omega^2\right) + g_{om}^2 \frac{\gamma_o}{2} (2n_{\text{as}}+1)}{(\omega-\omega_+^{pm})(\omega-\omega_-^{pm})(\omega-\omega_+^{pm\,*})(\omega-\omega_-^{pm\,*})}. 
    \end{aligned} \label{eq8}
\end{equation}
The final phonon occupation is obtained by integrating this displacement spectrum over frequency. This can be evaluated using contour integration in the lower half-plane and applying the residue theorem \cite{marquardt2007quantum} (see Appendix \ref{appC}). Combining with Eq.(\ref{eq6}), we obtain
\begin{equation}
    \begin{aligned}
        n_f = n_{\text{th}}\frac{\Gamma_m}{\gamma_o + \Gamma_m}\cdot \frac{4g_{om}^2 + \gamma_o(\gamma_o + \Gamma_m)}{4g_{om}^2 + \gamma_o\Gamma_m} + n_{\text{as}}\frac{\gamma_o}{\gamma_o + \Gamma_m}\cdot \frac{4g_{om}^2 }{4g_{om}^2 + \gamma_o\Gamma_m} + \frac{1}{2}. 
    \end{aligned} \label{eq9}
\end{equation}
Here, the first term coincides with the expression reported in Ref.\cite{blazquez2024oa} and represents the dominant contribution from the thermal bath, which is progressively suppressed as the optomechanical coupling increases. The remaining two terms set the fundamental limits of cooling through distinct quantum mechanisms. As the thermal influence is reduced, the relative importance of quantum backaction associated with the optical reservoir increases, as captured by the second term. This contribution originates from vacuum fluctuations of the anti-Stokes optical field and grows with increasing coupling strength. The third term, $1/2$, represents the zero-point motion of the acoustic oscillator, which persists even in the absence of thermal or optical excitation. Physically, reducing the thermal contribution does not eliminate the quantum fluctuations of the coupled system. Instead, optical vacuum fluctuations and intrinsic zero-point vibration become the dominant factors that set the cooling limit. 

\begin{figure}[h!]
  \includegraphics[width = 15cm]{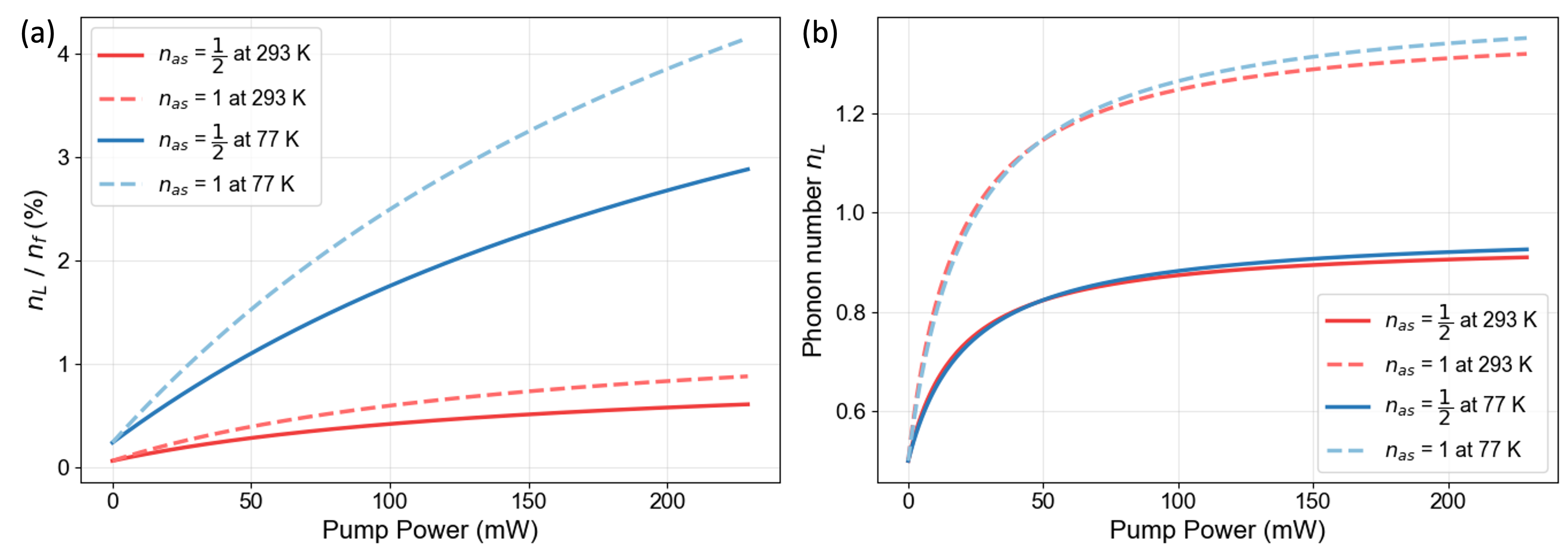}
  \caption{(a) Relative contribution of the quantum-limited term $n_L$ to the total phonon occupation $n_f$, and (b) quantum-limited phonon number $n_L$ as functions of pump power. Results are shown for room temperature (red) and cryogenic temperature $77~\mathrm{K}$ (blue), with $n_{\text{as}}=1$ (dashed lines) and $n_{\text{as}}=1/2$ (solid lines).} \label{figure4}
\end{figure}

For convenience, we group the last two terms and define $n_L$
\begin{equation}
    \begin{aligned}
        n_L = n_{\text{as}}\frac{\gamma_o}{\gamma_o + \Gamma_m}\cdot \frac{4g_{om}^2 }{4g_{om}^2 + \gamma_o\Gamma_m} + \frac{1}{2},
    \end{aligned} \label{eq10}
\end{equation}
which represents the quantum-limited contribution to the final phonon occupation. The ratio $n_L/n_f$ is shown in Fig.\ref{figure4}(a). Although $n_L$ is small compared to the thermal contribution at room temperature, it becomes increasingly significant in cryogenic environments (blue curves). The parameters at $77~\mathrm{K}$ are taken from Ref.\cite{fischer2025brillouin}, with $\gamma_o = 519$ MHz, $G_B = 196.1\mathrm{m^{-1}W^{-1}}$. $\Gamma_m = 36.8$ MHz is narrower than the 46.8 MHz measured at room temperature, which can be attributed to reduced acoustic damping at cryogenic temperatures suppressing other complicated dissipative processes, leading to an increased phonon lifetime. The core refractive index is assumed unchanged ($n = 2.5$). 

From the perspective of spontaneous emission, $n_{\text{as}}=1$ is often assumed, equivalent to a pseudo photon per mode in continuum system \cite{smith1972optical, Agrawal2013}. Alternatively, a value of $n_{\text{as}}=1/2$ can be adopted when considering purely quantum vacuum fluctuations. In the following, we examine both cases. As shown in Fig.\ref{figure4}(b), at low pump power ($P < 49$ mW), the quantum-limited contribution $n_L$ is weaker at cryogenic temperature than at room temperature. Once the pump power exceeds this value, the backaction in the cryogenic case becomes stronger. 

For $n_{\text{as}}=1$, the quantum backaction becomes comparable to zero-point fluctuations at a pump power of approximately $24$ mW at room temperature ($26$ mW at $77~\mathrm{K}$), preventing ground-state cooling. For $n_{\text{as}}=1/2$, the condition $n_L < 1$ can be satisfied within the undepleted pump approximation. However, achieving ground-state cooling remains challenging. Increasing the pump power enhances the cooling rate, but also amplifies quantum backaction. In the large pump-power limit, Eq.(\ref{eq9}) reduces to
\begin{equation}
    \begin{aligned}
        n_f \approx n_{\text{th}}\frac{\Gamma_m}{\gamma_o + \Gamma_m} + n_{\text{as}}\frac{\gamma_o}{\gamma_o + \Gamma_m} + \frac{1}{2}. 
    \end{aligned} \label{eq11}
\end{equation}
This expression illustrates a fundamental trade-off: while precooling reduces the thermal contribution by decreasing $\Gamma_m$, it simultaneously enhances the relative impact of quantum backaction, causing $n_L$ to approach unity. Even in the absence of quantum backaction ($n_{\text{as}}=0$), the irreducible zero-point fluctuations impose an extra lower bound on the phonon occupation, making the realization of ground-state cooling increasingly demanding.

Under the condition of $\gamma_o \gg \Gamma_m$, Eq.(\ref{eq9}) simplifies to
\begin{equation}
    \begin{aligned}
        n_f \approx n_{\text{th}} \frac{\Gamma_m}{\gamma_o}\cdot\frac{4g_{om}^2 + \gamma_o^2}{4g_{om}^2 + \gamma_o\Gamma_m} + n_{\text{as}}\frac{4g_{om}^2}{4g_{om}^2 + \gamma_o\Gamma_m} + \frac{1}{2}.
    \end{aligned} 
\end{equation}
The first two terms are formally analogous to the phonon occupation in cavity optomechanics under strong coupling when second-order perturbation are neglected \cite{aspelmeyer2014cavity, dobrindt2008parametric}. In cavity systems, such perturbation arise from the combined effect of Stokes and anti-Stokes processes acting on the same mechanical mode. In contrast, in waveguide optomechanics, these second-order processes are absent due to the symmetry breaking between Stokes and anti-Stokes processes.

\section{Conclusion}
We analyze that the traveling hypersound phonon cooling platform based on the tapered chalcogenide PCF can access the strong coupling regime at room temperature. The displacement spectrum of both anti-Stokes optical modes and acoustic modes are derived from the Lindblad equation for open quantum systems. Two key spectroscopic signatures of strong coupling, NMS and avoided crossing, emerge in the evolution of the full anti-Stokes spectrum as a function of pump power. By examining the eigenvalues of the effective non-Hermitian Hamiltonian, we determine that the threshold for strong coupling is at a relatively low pump peak power of approximately $27.3~\mathrm{mW}$. In addition, the theoretically predicted saturated threshold of the anti-Stokes spectral amplitude agrees with previously reported experimental observations. 

We further elucidate that achieving steady cooling of macroscopic phonons to ground-state in waveguide optomechanical systems is fundamentally limited by quantum backaction and zero-point fluctuations. While these effects are weak compared to thermal occupation at room temperature, they become increasingly significant in cryogenic environments. This prediction suggests that future experiments, for example in liquid helium cryostats operating at 4 K, could provide a direct test of the quantum limits identified in this work. These findings have important implications for accessing quantum behavior in waveguide optomechanical systems. Ground-state cooling ($n_f < 1$) is often considered a prerequisite for observing effects such as photon–phonon entanglement and quantum state transfer. Our results show that steady-state cooling alone is insufficient to realize these phenomena in the present platform, necessitating new paradigms beyond SBS-based laser-cooling approaches. In this sense, our work provides a revised framework for understanding the quantum limits of continuum optomechanical systems.

\begin{acknowledgments}
The authors thank Yuelang Huang and Christina Anna Louka for their helpful discussions. This work was supported by the Canada Research Chairs (950-231352) and Natural Sciences and Engineering Research Council of Canada (RGPIN-2020-06302, 06302/DGDND/2020).
\end{acknowledgments}

\appendix

\section{Dynamic Equations of Linearized Brillouin Interaction}
\label{appA}
Backward Brillouin anti-Stokes scattering can be formulated as an open quantum subsystem interacting with external environment. In this framework, the Schrödinger equation for pure states is generalized to the Lindblad master equation, which describes the dynamics of mixed states in terms of the density operator $\rho$ while preserving its trace and positivity \cite{breuer2002theory}. In the Schrödinger picture, the Lindblad equation reads
\begin{equation}
    \frac{d\rho}{dt} = -\frac{i}{\hbar}[\mathcal{H}, \rho] + \sum_j \left( L_j \rho L_j^{\dagger} - \frac{1}{2} \{L_j^{\dagger}L_j, \rho \}\right), 
\end{equation}
where $L_j$ are the Lindblad (jump) operators that determines the dissipative component of the system’s evolution \cite{nathan2020universal}. The equivalent expression for an arbitrary system operator $\mathcal{O}$ in the Heisenberg picture is given by
\begin{equation}
    \frac{d\mathcal{O}}{dt} = \frac{i}{\hbar} [\mathcal{H}, \mathcal{O}] + \sum_j\left(L_j^{\dagger}\mathcal{O} L_j - \frac{1}{2} L_j^{\dagger}L_j \mathcal{O} - \frac{1}{2} \mathcal{O} L_j^{\dagger}L_j \right). \label{eqa2}
\end{equation}
The first term describes the unitary evolution generated by the Hamiltonian $\mathcal{H}$, while the remaining terms account for dissipation due to coupling with the environment. The total Hamiltonian can be decomposed as
\begin{equation}
    \mathcal{H} = \mathcal{H}_{S} + \mathcal{H}_{B} + \mathcal{H}_{int}, 
\end{equation}
where $\mathcal{H}_{S}$, $\mathcal{H}_{B}$, and $\mathcal{H}_{int}$ describe the subsystem, the bath, and their interaction, respectively. 

Under the undepleted pump approximation, the linearized Hamiltonian for backward Brillouin anti-Stokes scattering is given by \cite{blazquez2024oa}
\begin{equation}
    \mathcal{H}_S = \int \hbar\omega_{as}(k) a^{\dagger}_{as}(k) a_{as}(k)dk + \int \hbar\Omega_{ac}(q) b^{\dagger}_{ac}(q) b_{ac}(q)dq + \hbar g_{om} \iint a^{\dagger}_{as}(k) b_{ac}(q)dkdq + h.c.,
\end{equation}
where $a_{as}(k,t)$ and $b_{ac}(q,t)$ represent the envelope operators of the anti-Stokes and acoustic mode with wavenumber $k$ and $q$, respectively. $g_{om}=g_0\sqrt{\langle a_p^{\dagger} a_p\rangle}$ is the effective pump-enhanced coupling strength, where $g_0$ is the vacuum coupling strength which quantifies the interaction intensity between a single photon and a single phonon \cite{tomasella2025strong, zhu2023dynamic, van2016unifying}. The bath Hamiltonian and the system–bath interaction Hamiltonian take the form
\begin{equation}
    \begin{aligned}
        &\mathcal{H}_B = \int \hbar\omega_{as}(k) \xi^{\dagger}_{as}(k) \xi_{as}(k)dk + \int \hbar\Omega_{ac}(q) \xi^{\dagger}_{ac}(q) \xi_{ac}(q)dq,\\[5pt]
        &\mathcal{H}_{int} = i\hbar\int \left[\xi^{\dagger}_{as}(k)L_a + L_a^{\dagger}\xi_{as}(k)\right]dk + i\hbar\int \left[\xi^{\dagger}_{ac}(q)L_b + L_b^{\dagger}\xi_{ac}(q)\right]dq, 
    \end{aligned} 
\end{equation}
where $\xi_{as}(k)$ and $\xi_{ac}(q)$ denote the quantum noise operators associated with the optical and acoustic reservoirs, and satisfy the following statistical properties 
\begin{equation}
    \begin{aligned}
        &\left\langle \xi_{as}(k,t) \right\rangle = \left\langle \xi_{ac}(q,t) \right\rangle = 0, \\[5pt]
        &\left\langle \xi_{as}^{\dagger}(k_1,t_1) \xi_{as}(k_2,t_2) \right\rangle = n_{\text{as}}\delta(k_1 - k_2)\delta(t_1 - t_2), \\[5pt]
        &\left\langle \xi_{ac}^{\dagger}(q_1,t_1) \xi_{ac}(q_2,t_2) \right\rangle = n_{\text{th}}\delta(q_1 - q_2)\delta(t_1 - t_2),
    \end{aligned} \label{eqa6}
\end{equation} 
where $n_{\text{as}}$ is defined through the noise correlation and $n_{\text{th}} = \left[\text{exp}(\hbar \Omega_{ac}/k_BT) - 1\right]^{-1}$ is the thermal phonon occupation at temperature $T$. The number of jump operators equals the number of independent quantum noise channels \cite{nathan2020universal}. Accordingly, we introduce $L_a = \sqrt{\gamma_o}, a_{as}(k)$ and $L_b = \sqrt{\Gamma_m}, b_{ac}(q)$, where $\gamma_o$ and $\Gamma_m$ are the optical and acoustic damping rates, respectively. 

Substituting the total Hamiltonian $\mathcal{H}$ into the Eq.(\ref{eqa2}), we get
\begin{equation}
    \begin{aligned}
        \frac{d a_{as}(k')}{dt} =&\ i \int dk\ \omega_{as}(k)a_{as}(k) \left[a^{\dagger}_{as}(k),\ a_{as}(k') \right] + ig_{om} \iint dkdq\  b_{ac}(q) \left[a^{\dagger}_{as}(k),\ a_{as}(k') \right] \\[5pt] 
        & + \left(\frac{1}{2}\gamma_o a_{as}^{\dagger}(k)a_{as}(k')a_{as}(k) - \frac{1}{2}\gamma_o a_{as}^{\dagger}(k)a_{as}(k)a_{as}(k') - \frac{1}{2} \gamma_o a_{as}(k)\right) \\[5pt]
        & - \int dk\ \left[\xi^{\dagger}_{as}(k)L_a, a_{as}(k')\right] - \int dk\ \left[L_a^{\dagger}\xi_{as}(k), a_{as}(k')\right]. \\[5pt]
    \end{aligned}
\end{equation}
Due to the commutation relation $\left[\xi^{\dagger}_{as}(k)L_a, a_{as}(k')\right]=0$, we have
\begin{equation}
    \begin{aligned}
        \frac{d a_{as}(k')}{dt} =&\ i \int dk\ \omega_{as}(k)a_{as}(k) \left[a^{\dagger}_{as}(k),\ a_{as}(k') \right] + ig_{om} \iint dkdq\  b_{ac}(q) \left[a^{\dagger}_{as}(k),\ a_{as}(k') \right] \\[5pt] 
        & + \left(\frac{1}{2}\gamma_o a_{as}^{\dagger}(k)a_{as}(k')a_{as}(k) - \frac{1}{2}\gamma_o a_{as}^{\dagger}(k)a_{as}(k)a_{as}(k') - \frac{1}{2} \gamma_o a_{as}(k)\right) \\[5pt]
        & - \int dk\ \left[\sqrt{\gamma_o}a_{as}^{\dagger}(k), a_{as}(k')\right]\xi_{as}(k),
    \end{aligned}
\end{equation}
or
\begin{equation}
    \frac{d a_{as}(k)}{dt} = -i\omega_{as}(k) a_{as}(k) - ig_{om} b_{ac}(q) - \frac{\gamma_o}{2}a_{as}(k) + \sqrt{\gamma_o} \xi_{as}(k). 
\end{equation}
A similar procedure applied to the operator $b_{ac}(q,t)$ yields
\begin{equation}
    \frac{d b_{ac}(q)}{dt} = -i\Omega_{ac}(q) b_{ac}(q) - ig_{om} a_{as}(k) -\frac{\Gamma_m}{2}b_{ac}(q) + \sqrt{\Gamma_m} \xi_{ac}(q). 
\end{equation}
By using the method of frame rotating,\cite{blazquez2024oa} the dynamics of the linearized anti-Stokes scattering process can be rewritten as
\begin{subequations}
    \label{eqa11}
    \begin{equation}
        \frac{d a(k,t)}{d t} = \left(-\frac{\gamma_o}{2} + i\Delta_{as}\right)a(k,t) -ig_{om} b(q,t) + \sqrt{\gamma_o}\xi_{a}(k,t), \label{eqa11a}
    \end{equation}
    \begin{equation}
        \frac{d b(q,t)}{d t} = \left(-\frac{\Gamma_m}{2} + i\Delta_{ac}\right)b(q,t) -ig_{om} a(k,t) + \sqrt{\Gamma_m}\xi_{b}(q,t), \label{eqa11b}
    \end{equation}
\end{subequations}
where $\Delta_{as} = kv_{o}$ and $\Delta_{ac} = qv_{ac}$ imply the frequency shifts induced by phase mismatching for the $k$-th anti-Stokes mode and $q$-th acoustic mode, respectively. $v_{o}$ ($v_{ac}$) denotes the group velocities of the anti-Stokes (acoustic) mode. For the cooling of additional acoustic modes supported in a continuum system, the frequency shifts induced by group velocity dispersion must also be taken into account. In particular, it becomes increasingly important for optical frequencies far from 1550 nm and in waveguide platforms with reduced core size \cite{yang2025optical}.

\section{Displacement Spectrum}
\label{appB}
Applying the Fourier transform to Eq.(\ref{eqa11}) and moving to the frequency domain, we obtain
\begin{subequations}
    \label{eqb1}
    \begin{equation}
        \left[\frac{\gamma_o}{2} - i(\omega + \Delta_{as})\right]\tilde{a}(k,\omega) = -ig_{om} \tilde{b}(q,\omega) + \sqrt{\gamma_o}\tilde{\xi}_a(k,\omega), \label{eqb1a}
    \end{equation}
    \begin{equation}
        \left[\frac{\Gamma_m}{2} - i(\omega + \Delta_{ac})\right]\tilde{b}(q,\omega) = -ig_{om} \tilde{a}(k,\omega) + \sqrt{\Gamma_m}\tilde{\xi}_b(q,\omega). \label{eqb1b}
    \end{equation}
\end{subequations}
Solving Eq.(\ref{eqb1}) yields
\begin{subequations}
    \label{eqb2}
    \begin{equation}
        \tilde{a}(k,\omega) = \frac{\left[\frac{\Gamma_m}{2} - i(\omega + \Delta_{ac})\right] \sqrt{\gamma_o}\tilde{\xi}_a(k,\omega) - ig_{om}\sqrt{\Gamma_m}\tilde{\xi}_b(q,\omega)}{g_{om}^2 + \left[\frac{\gamma_o}{2} - i(\omega + \Delta_{as})\right]\left[\frac{\Gamma_m}{2} - i(\omega + \Delta_{ac})\right]}, \label{eqb2a}
    \end{equation}
    \begin{equation}
        \tilde{b}(q,\omega) = \frac{\left[\frac{\gamma_o}{2} - i(\omega + \Delta_{as})\right] \sqrt{\Gamma_m}\tilde{\xi}_b(q,\omega) - ig_{om}\sqrt{\gamma_o}\tilde{\xi}_a(k,\omega)}{g_{om}^2 + \left[\frac{\gamma_o}{2} - i(\omega + \Delta_{as})\right]\left[\frac{\Gamma_m}{2} - i(\omega + \Delta_{ac})\right]}. \label{eqb2b}
    \end{equation}
\end{subequations}
Based on Eq.(\ref{eqa6}), the Langevin noise operators in frequency domain satisfy the following correlation relations
\begin{equation}
    \begin{aligned}
        &\left\langle \tilde{\xi}_a^{\dagger}(k,\omega_1) \tilde{\xi}_a(k,\omega_2) \right\rangle = 2\pi n_{\text{as}}\delta(\omega_1 - \omega_2), \qquad
        \left\langle \tilde{\xi}_a(k,\omega_1) \tilde{\xi}_a^{\dagger}(k,\omega_2) \right\rangle = 2\pi(n_{\text{as}}+1)\delta(\omega_1 - \omega_2); \\[5pt]
        &\left\langle \tilde{\xi}_b^{\dagger}(q,\omega_1) \tilde{\xi}_b(q,\omega_2) \right\rangle = 2\pi n_{\text{th}}\delta(\omega_1 - \omega_2), \qquad
        \left\langle \tilde{\xi}_b(q,\omega_1) \tilde{\xi}_b^{\dagger}(q,\omega_2) \right\rangle = 2\pi(n_{\text{th}}+1)\delta(\omega_1 - \omega_2). \\[5pt]
    \end{aligned} \label{eqb3}
\end{equation}

The quadrature operators associated with the anti-Stokes and acoustic modes are defined as
\begin{equation}
    \begin{aligned}
        &X_a(k,t) = \frac{a(k,t) + a^{\dagger}(k,t)}{\sqrt{2}}, \quad Y_a(k,t) = i\frac{a(k,t) - a^{\dagger}(k,t)}{\sqrt{2}}; \\[5pt]
        &X_b(q,t) = \frac{b(q,t) + b^{\dagger}(q,t)}{\sqrt{2}}, \quad Y_b(q,t) = i\frac{b(q,t) - b^{\dagger}(q,t)}{\sqrt{2}}. 
    \end{aligned} \label{eqb4}
\end{equation}
The displacement spectrum $S_{XX}^a(k,\omega)$ of the $k$th anti-Stokes mode is obtained from the relation \cite{agarwal2013multimode, chen2016brillouin}
\begin{equation}
    \begin{aligned}
        \left\langle \tilde{X}_a^{\dagger}(k,\omega_1)\tilde{X}_a(k,\omega_2)\right\rangle =&\ \frac{1}{2}\left[\left\langle \tilde{a}^{\dagger}(k,\omega_1) \tilde{a}(k,\omega_2)\right\rangle + \left\langle \tilde{a}(k,\omega_1) \tilde{a}^{\dagger}(k,\omega_2)\right\rangle\right] \\[5pt]
        &+ \frac{1}{2}\left[\left\langle \tilde{a}^{\dagger}(k,\omega_1) \tilde{a}^{\dagger}(k,\omega_2)\right\rangle + \left\langle \tilde{a}(k,\omega_1) \tilde{a}(k,\omega_2)\right\rangle\right] \\[5pt]
        =&\ 2\pi S_{XX}^a(k, \omega_1)\delta(\omega_1 - \omega_2). 
    \end{aligned} \label{eqb5}
\end{equation}
Using the solution for the anti-Stokes mode in Eq.(\ref{eqb2a}) together with the noise correlations in Eq.(\ref{eqb3}), we get
\begin{equation*}
\begin{aligned}
    \begin{split}
        &\left\langle \tilde{a}^{\dagger}(k,\omega_1)\tilde{a}(k,\omega_2)\right\rangle \\[5pt]
        =&\ \Bigg\langle
        \frac{\left[\frac{\Gamma_m}{2} + i(\omega_1 + \Delta_{ac})\right]\sqrt{\gamma_o}\tilde{\xi}^{\dagger}_a(k,\omega_1) + i g_{om}\sqrt{\Gamma_m}\tilde{\xi}_b^{\dagger}(q,\omega_1)}
        {g_{om}^2 + \left[\frac{\gamma_o}{2} + i(\omega_1 + \Delta_{as})\right] \left[\frac{\Gamma_m}{2} + i(\omega_1 + \Delta_{ac})\right]} \\[5pt]
        &\ \times
        \frac{\left[\frac{\Gamma_m}{2} - i(\omega_2 + \Delta_{ac})\right]\sqrt{\gamma_o}\tilde{\xi}_a(k,\omega_2)
        - i g_{om}\sqrt{\Gamma_m}\tilde{\xi}_b(q,\omega_2)}
        {g_{om}^2 + \left[\frac{\gamma_o}{2} - i(\omega_2 + \Delta_{as})\right]
        \left[\frac{\Gamma_m}{2} - i(\omega_2 + \Delta_{ac})\right]}
        \Bigg\rangle \\[5pt]
        &=\ \frac{\gamma_o \left[\frac{\Gamma_m}{2} + i(\omega_1 + \Delta_{ac})\right] \left[\frac{\Gamma_m}{2} - i(\omega_2 + \Delta_{ac})\right] \left\langle \tilde{\xi}_a^{\dagger}(k,\omega_1) \tilde{\xi}_a(k,\omega_2) \right\rangle + g_{om}^2 \Gamma_m \left\langle \tilde{\xi}_b^{\dagger}(q,\omega_1) \tilde{\xi}_b(q,\omega_2) \right\rangle}{\left\{g_{om}^2 + \left[\frac{\gamma_o}{2} + i(\omega_1 + \Delta_{as})\right]\left[\frac{\Gamma_m}{2} + i(\omega_1 + \Delta_{ac})\right] \right\} \left\{ g_{om}^2 + \left[\frac{\gamma_o}{2} - i(\omega_2 + \Delta_{as})\right]\left[\frac{\Gamma_m}{2} - i(\omega_2 + \Delta_{ac})\right]\right\}}, 
    \end{split}
\end{aligned}
\end{equation*}
or
\begin{equation}
    \begin{aligned}
        &\left\langle \tilde{a}^{\dagger}(k,\omega_1) \tilde{a}(k,\omega_2)\right\rangle \\[5pt] =&\ 2\pi\frac{\left\{\gamma_o n_{\text{as}} \left[\frac{\Gamma_m}{2} + i(\omega_1 + \Delta_{ac})\right] \left[\frac{\Gamma_m}{2} - i(\omega_2 + \Delta_{ac})\right] + g_{om}^2 \Gamma_m n_{\text{th}} \right\} \delta(\omega_1 - \omega_2)}{\left\{g_{om}^2 + \left[\frac{\gamma_o}{2} + i(\omega_1 + \Delta_{as})\right]\left[\frac{\Gamma_m}{2} + i(\omega_1 + \Delta_{ac})\right] \right\} \left\{ g_{om}^2 + \left[\frac{\gamma_o}{2} - i(\omega_2 + \Delta_{as})\right]\left[\frac{\Gamma_m}{2} - i(\omega_2 + \Delta_{ac})\right]\right\}}.
    \end{aligned} \label{eqb6}
\end{equation}
An analogous derivation leads to
\begin{equation}
    \begin{aligned}
        &\left\langle \tilde{a}(k,\omega_1) \tilde{a}^{\dagger}(k,\omega_2)\right\rangle \\[5pt] 
        &= 2\pi \frac{\left\{\gamma_o (n_{\text{as}} + 1) \left[\frac{\Gamma_m}{2} - i(\omega_1 + \Delta_{ac})\right] \left[\frac{\Gamma_m}{2} + i(\omega_2 + \Delta_{ac})\right] + g_{om}^2 \Gamma_m (n_{\text{th}} + 1)\right\} \delta(\omega_1 - \omega_2)}{\left\{g_{om}^2 + \left[\frac{\gamma_o}{2} - i(\omega_1 + \Delta_{as})\right]\left[\frac{\Gamma_m}{2} - i(\omega_1 + \Delta_{ac})\right] \right\} \left\{ g_{om}^2 + \left[\frac{\gamma_o}{2} + i(\omega_2 + \Delta_{as})\right]\left[\frac{\Gamma_m}{2} + i(\omega_2 + \Delta_{ac})\right]\right\}}, \\[5pt]
        &\left\langle \tilde{a}^{\dagger}(k,\omega_1) \tilde{a}^{\dagger}(k,\omega_2)\right\rangle = 0, \\[5pt]
        &\left\langle \tilde{a}(k,\omega_1) \tilde{a}(k,\omega_2)\right\rangle = 0.
    \end{aligned} \label{eqb7}
\end{equation}
Combining Eq.(\ref{eqb6}) and Eq.(\ref{eqb7}) with Eq.(\ref{eqb5}), the displacement spectrum of the $k$th anti-Stokes mode is obtained as
\begin{equation}
    \begin{aligned}
        S_{XX}^a(k, \omega) = \frac{\frac{\gamma_o}{2} (2n_{\text{as}} + 1) \left[\frac{\Gamma_m^2}{4} + (\omega + \Delta_{ac})^2\right] + g_{om}^2 \frac{\Gamma_m}{2} (2n_{\text{th}}+1)}{\left| g_{om}^2 + \left[\frac{\gamma_o}{2} - i(\omega + \Delta_{as})\right]\left[\frac{\Gamma_m}{2} - i(\omega + \Delta_{ac})\right] \right|^2}.
    \end{aligned} \label{eqb8}
\end{equation}
We introduce the response functions of the optical and acoustic mode as $\chi_{as}(\omega) = \left[\frac{\gamma_o}{2} - i(\omega + \Delta_{as})\right]^{-1}$ and $\chi_{ac}(\omega) = \left[\frac{\Gamma_m}{2} - i(\omega + \Delta_{ac})\right]^{-1}$. Defining $D(\omega)=g_{om}^2 + \chi_{as}^{-1}(\omega) \chi_{ac}^{-1}(\omega)$, the displacement spectrum can be rewritten as
\begin{equation}
    S_{XX}^a(k, \omega) = \frac{\frac{\gamma_o}{2}(2n_{\text{as}} + 1) \left[\frac{\Gamma_m^2}{4} + (\omega + \Delta_{ac})^2\right] + g_{om}^2 \frac{\Gamma_m}{2} (2n_{\text{th}}+1)}{|D(\omega)|^2}. \label{eqb9}
\end{equation}
Similarly, the displacement spectrum of the $q$-th acoustic mode is given by
\begin{equation}
    \begin{aligned}
        S_{XX}^b(q, \omega) &= \frac{\frac{\Gamma_m}{2} (2n_{\text{th}}+1) \left[\frac{\gamma_o^2}{4} + (\omega + \Delta_{as})^2\right] + g_{om}^2 \frac{\gamma_o}{2} (2n_{\text{as}} + 1)}{\left| g_{om}^2 + \left[\frac{\gamma_o}{2} - i(\omega + \Delta_{as})\right]\left[\frac{\Gamma_m}{2} - i(\omega + \Delta_{ac})\right] \right|^2} \\[5pt]
        &= \frac{\frac{\Gamma_m}{2} (2n_{\text{th}}+1) \left[\frac{\gamma_o^2}{4} + (\omega + \Delta_{as})^2\right] + g_{om}^2 \frac{\gamma_o}{2} (2n_{\text{as}} + 1)}{|D(\omega)|^2}.
    \end{aligned} \label{eqb10}
\end{equation}

\section{Phonon Occupation in Waveguide Optomechanics}
\label{appC}
Eq.(\ref{eqa11}) can be re-expressed in matrix form as 
\begin{equation}
    \frac{d}{dt} \begin{pmatrix}
        a \\ b
    \end{pmatrix} = -i\begin{pmatrix}
        -\Delta_{as} - i\frac{\gamma_o}{2} & g_{om} \\
        g_{om} & -\Delta_{ac} - i\frac{\Gamma_m}{2}
    \end{pmatrix} \begin{pmatrix}
        a \\ b
    \end{pmatrix} + \begin{pmatrix}
        \sqrt{\gamma_o}\xi_{a} \\ \sqrt{\Gamma_m}\xi_{b}
    \end{pmatrix}. 
\end{equation}
Since the Langevin noise has zero mean, $\langle \xi_i \rangle = 0$ ($i = a,b$), it does not affect the intrinsic eigenfrequencies or decay rates of the system. The complex eigenfrequencies are therefore obtained from the homogeneous part of the equations of motion as
\begin{equation}
    \omega_{\pm} = -\frac{\Delta_{as} + \Delta_{ac}}{2} -i\frac{\gamma_o + \Gamma_m}{4} \pm \sqrt{g_{om}^2 + \left(\frac{\Delta_{as} - \Delta_{ac}}{2} + i\frac{\gamma_o - \Gamma_m}{4}\right)^2}, \label{eqc2}
\end{equation}
which are also the solutions of $D(\omega)=0$. Accordingly, $D(\omega)$ and its conjugate can be factorized as $D(\omega)= -(\omega-\omega_+)(\omega-\omega_-)$ and $D^*(\omega)= -(\omega-\omega_+^*)(\omega-\omega_-^*)$. Therefore, the denominator of the displacement spectrum can be expressed as
\begin{equation}
    |D(\omega)|^2=(\omega-\omega_+)(\omega-\omega_-)(\omega-\omega_+^*)(\omega-\omega_-^*), \label{eqc3}
\end{equation}
which means that $S_{XX}^a(k, \omega)$ and $S_{XX}^b(q, \omega)$ possess four poles, $\omega_{\pm}$ in the lower half of the complex plane and $\omega_{\pm}^*$ in the upper half plane. For the perfect phase-matched interaction, the eigenvalues of system is reduced to 
\begin{equation}
    \omega_{\pm}^{pm} = -i\frac{\gamma_o + \Gamma_m}{4} \pm \sqrt{g_{om}^2 - \left(\frac{\gamma_o - \Gamma_m}{4}\right)^2}.  \label{eqc4}
\end{equation}
By combining Eq.(\ref{eqc3}) and Eq.(\ref{eqc4}), the displacement spectrum for the phase-matched acoustic mode is expressed as
\begin{equation}
    \begin{aligned}
        S_{XX}^b(q=0, \omega) &= \frac{\frac{\Gamma_m}{2} (2n_{\text{th}}+1) \left(\frac{\gamma_o^2}{4} + \omega^2\right) + g_{om}^2 \frac{\gamma_o}{2} (2n_{\text{as}} + 1)}{|D(\omega)|^2} \\[5pt] &= \frac{\frac{\Gamma_m}{2} (2n_{\text{th}}+1) \left(\frac{\gamma_o^2}{4} + \omega^2\right) + g_{om}^2 \frac{\gamma_o}{2} (2n_{\text{as}} + 1)}{(\omega-\omega_+^{pm})(\omega-\omega_-^{pm})(\omega-\omega_+^{pm\,*})(\omega-\omega_-^{pm\,*})}. 
    \end{aligned} \label{eqc5}
\end{equation}

We now perform contour integration on $S_{XX}^b(q=0, \omega)$ in the lower half plane and use the residue theorem \cite{marquardt2007quantum}
\begin{equation}
    \begin{aligned}
    \begin{split}
        n_f =&\ \frac{1}{2\pi}\int d\omega\ S^b_{XX}(q=0,\omega) \\[5pt] =&\ \frac{1}{2\pi} \cdot (-2\pi i)\sum \text{Res}\left[S^b_{XX}(q=0,\omega) \right]_{\omega = \omega_{\pm}^{pm}} \\[5pt]
        =&\ -i\Bigg[\frac{\frac{\Gamma_m}{2} (2n_{\text{th}}+1) \left(\frac{\gamma_o^2}{4} + \omega_+^{pm\,2}\right) + g_{om}^2 \frac{\gamma_o}{2} (2n_{\text{as}} + 1)}{(\omega_+^{pm}-\omega_-^{pm})(\omega_+^{pm}-\omega_+^{pm\,*})(\omega_+^{pm}-\omega_-^{pm\,*})} \\[5pt]
        &\qquad \ \ + \frac{\frac{\Gamma_m}{2} (2n_{\text{th}}+1) \left(\frac{\gamma_o^2}{4} + \omega_-^{pm\,2}\right) + g_{om}^2 \frac{\gamma_o}{2} (2n_{\text{as}} + 1)}{(\omega_-^{pm}-\omega_+^{pm})(\omega_-^{pm}-\omega_+^{pm\,*})(\omega_-^{pm}-\omega_-^{pm\,*})}\Bigg]. 
    \end{split}
    \end{aligned} \label{eqc6}
\end{equation}
Using Eq.(\ref{eqc4}), we find
\begin{equation}
    \begin{aligned}
        \frac{\gamma_o^2}{4} + \omega_+^{pm\,2} &= \frac{\gamma_o^2}{4} + \left[-i\frac{\gamma_o + \Gamma_m}{4} + \sqrt{g_{om}^2 - \left(\frac{\gamma_o - \Gamma_m}{4}\right)^2}\right]^2 \\[5pt]
        &= g_{om}^2 + \frac{1}{8}(\gamma_o + \Gamma_m)(\gamma_o - \Gamma_m) - i \frac{\gamma_o + \Gamma_m}{2} \sqrt{g_{om}^2 - \left(\frac{\gamma_o - \Gamma_m}{4}\right)^2},
    \end{aligned}
\end{equation}
and
\begin{equation}
    \begin{aligned}
        &(\omega_+^{pm}-\omega_-^{pm}) = 2\sqrt{g_{om}^2 - \left(\frac{\gamma_o - \Gamma_m}{4}\right)^2}, \qquad (\omega_+^{pm}-\omega_+^{pm\,*}) = -i\frac{\gamma_o + \Gamma_m}{2}, \\[5pt] 
        &(\omega_+^{pm}-\omega_-^{pm\,*}) = -i\frac{\gamma_o + \Gamma_m}{2} + 2\sqrt{g_{om}^2 - \left(\frac{\gamma_o - \Gamma_m}{4}\right)^2}.
    \end{aligned}
\end{equation}
The first term of Eq.(\ref{eqc6}) would thus be
\begin{equation}
    -i\frac{\frac{\Gamma_m}{2} (2n_{\text{th}}+1) \left[g_{om}^2 + \frac{1}{8}(\gamma_o + \Gamma_m)(\gamma_o - \Gamma_m) - i \frac{\gamma_o + \Gamma_m}{2} \sqrt{g_{om}^2 - \left(\frac{\gamma_o - \Gamma_m}{4}\right)^2} \right] + g_{om}^2 \frac{\gamma_o}{2} (2n_{\text{as}} + 1)}{-\frac{(\gamma_o + \Gamma_m)^2}{2} \sqrt{g_{om}^2 - \left(\frac{\gamma_o - \Gamma_m}{4}\right)^2} -i2(\gamma_o + \Gamma_m) \left[g_{om}^2 - \left(\frac{\gamma_o - \Gamma_m}{4}\right)^2\right]}.
\end{equation}
By the same procedure, the second term of Eq.(\ref{eqc6}) can be written as
\begin{equation}
    -i\frac{\frac{\Gamma_m}{2} (2n_{\text{th}}+1) \left[ g_{om}^2 + \frac{1}{8}(\gamma_o + \Gamma_m)(\gamma_o - \Gamma_m) + i \frac{\gamma_o + \Gamma_m}{2} \sqrt{g_{om}^2 - \left(\frac{\gamma_o - \Gamma_m}{4}\right)^2}\right] + g_{om}^2 \frac{\gamma_o}{2} (2n_{\text{as}} + 1)}{\frac{(\gamma_o + \Gamma_m)^2}{2} \sqrt{g_{om}^2 - \left(\frac{\gamma_o - \Gamma_m}{4}\right)^2} -i2(\gamma_o + \Gamma_m) \left[g_{om}^2 - \left(\frac{\gamma_o - \Gamma_m}{4}\right)^2\right]}.
\end{equation}
The sum of these contributions then yields
\begin{equation}
    \begin{aligned}
        n_f &= \frac{2\Gamma_m (2n_{\text{th}}+1)\left[g_{om}^2 +\frac{1}{4} \gamma_o(\gamma_o + \Gamma_m)\right] + 2g_{om}^2 \gamma_o (2n_{\text{as}} + 1)}{4(\gamma_o + \Gamma_m)\left[g_{om}^2 - \left(\frac{\gamma_o - \Gamma_m}{4}\right)^2\right] + \frac{(\gamma_o + \Gamma_m)^3}{4}} \\[5pt]
        &= n_{\text{th}}\frac{4\Gamma_m \left[g_{om}^2 +\frac{1}{4} \gamma_o(\gamma_o + \Gamma_m)\right]}{(4g_{om}^2 + \gamma_o\Gamma_m)(\gamma_o + \Gamma_m)} 
        + \frac{2\Gamma_m \left[g_{om}^2 +\frac{1}{4} \gamma_o(\gamma_o + \Gamma_m)\right]}{(4g_{om}^2 + \gamma_o\Gamma_m)(\gamma_o + \Gamma_m)} 
        + \frac{2g_{om}^2 \gamma_o (2n_{\text{as}} + 1)}{(4g_{om}^2 + \gamma_o\Gamma_m)(\gamma_o + \Gamma_m)}\\[10pt]
        &= n_{\text{th}}\frac{\Gamma_m}{\gamma_o + \Gamma_m} \cdot \frac{4g_{om}^2 + \gamma_o(\gamma_o + \Gamma_m)}{4g_{om}^2 + \gamma_o\Gamma_m} + n_{\text{as}}\frac{\gamma_o}{\gamma_o + \Gamma_m}\cdot \frac{4g_{om}^2}{4g_{om}^2 + \gamma_o\Gamma_m} + \frac{1}{2}. 
    \end{aligned} \label{eqc11}
\end{equation}

\nocite{*}

\bibliography{apssamp}

@article{whittle2021approaching,
  title={Approaching the motional ground state of a 10-kg object},
  author={Whittle, Chris and Hall, Evan D and Dwyer, Sheila and Mavalvala, Nergis and Sudhir, Vivishek and Abbott, Robert and Ananyeva, A and Austin, C and Barsotti, L and Betzwieser, Joseph and others},
  journal={Science},
  volume={372},
  number={6548},
  pages={1333--1336},
  year={2021},
  publisher={American Association for the Advancement of Science}
}

@article{gras2017audio,
  title={Audio-band coating thermal noise measurement for Advanced {LIGO} with a multimode optical resonator},
  author={Gras, Slawek and Yu, H and Yam, W and Martynov, Denis and Evans, M},
  journal={Physical Review D},
  volume={95},
  number={2},
  pages={022001},
  year={2017},
  publisher={APS}
}

@article{abbott2009observation,
  title={Observation of a kilogram-scale oscillator near its quantum ground state},
  author={Abbott, B and Abbott, R and Adhikari, R and Ajith, Parameswaran and Allen, Bruce and Allen, G and Amin, R and Anderson, SB and Anderson, WG and Arain, MA and others},
  journal={New Journal of Physics},
  volume={11},
  number={7},
  pages={073032},
  year={2009}
}

@article{gonzalez2000suspensions,
  title={Suspensions thermal noise in the {LIGO} gravitational wave detector},
  author={Gonz{\'a}lez, Gabriela},
  journal={Classical and Quantum Gravity},
  volume={17},
  number={21},
  pages={4409--4435},
  year={2000}
}

@article{bize2005cold,
  title={Cold atom clocks and applications},
  author={Bize, S{\'e}bastien and Laurent, Philippe and Abgrall, Michel and Marion, Harold and Maksimovic, Ivan and Cacciapuoti, Luigi and Gr{\"u}nert, Jan and Vian, C{\'e}line and Santos, F Pereira dos and Rosenbusch, Peter and others},
  journal={Journal of Physics B: Atomic, Molecular and Optical physics},
  volume={38},
  number={9},
  pages={S449--S468},
  year={2005}
}

@article{kessler2012sub,
  title={A sub-40-{mHz}-linewidth laser based on a silicon single-crystal optical cavity},
  author={Kessler, Thomas and Hagemann, Christian and Grebing, C and Legero, T and Sterr, Uwe and Riehle, Fritz and Martin, MJ and Chen, L and Ye, J},
  journal={Nature Photonics},
  volume={6},
  number={10},
  pages={687--692},
  year={2012},
  publisher={Nature Publishing Group UK London}
}

@article{ludlow2015optical,
  title={Optical atomic clocks},
  author={Ludlow, Andrew D and Boyd, Martin M and Ye, Jun and Peik, Ekkehard and Schmidt, Piet O},
  journal={Reviews of Modern Physics},
  volume={87},
  number={2},
  pages={637--701},
  year={2015},
  publisher={APS}
}

@article{liu2018orbit,
  title={In-orbit operation of an atomic clock based on laser-cooled {$^{87}$Rb} atoms},
  author={Liu, Liang and L{\"u}, De-Sheng and Chen, Wei-Biao and Li, Tang and Qu, Qiu-Zhi and Wang, Bin and Li, Lin and Ren, Wei and Dong, Zuo-Ren and Zhao, Jian-Bo and others},
  journal={Nature Communications},
  volume={9},
  number={1},
  pages={2760},
  year={2018},
  publisher={Nature Publishing Group UK London}
}

@article{Reilly2026fully,
  title={Fully collective superradiant lasing with vanishing sensitivity to cavity length vibrations},
  author={Reilly, Jarrod T and J$\ddot{a}$ger, Simon B and Cooper, John and Holland, Murray J},
  journal={Physical Review Letters},
  volume={136},
  number={14},
  pages={143803},
  year={2026},
  publisher={APS}
}

@article{odeh2025non,
  title={Non-Markovian dynamics of a superconducting qubit in a phononic bandgap},
  author={Odeh, Mutasem and Godeneli, Kadircan and Li, Eric and Tangirala, Rohin and Zhou, Haoxin and Zhang, Xueyue and Zhang, Zi-Huai and Sipahigil, Alp},
  journal={Nature Physics},
  volume={21},
  number={3},
  pages={406--411},
  year={2025},
  publisher={Nature Publishing Group UK London}
}

@article{bassman2024dynamic,
  title={Dynamic cooling on contemporary quantum computers},
  author={Bassman Oftelie, Lindsay and De Pasquale, Antonella and Campisi, Michele},
  journal={PRX Quantum},
  volume={5},
  number={3},
  pages={030309},
  year={2024},
  publisher={APS}
}

@article{fabrikant2024cooling,
  title={Cooling trapped ions with phonon rapid adiabatic passage},
  author={Fabrikant, MI and Lauria, P and Madjarov, IS and Burton, WC and Sutherland, RT},
  journal={Physical Review X},
  volume={14},
  number={4},
  pages={041046},
  year={2024},
  publisher={APS}
}

@article{feng2022quant,
  title={Quantum Computing by Coherent Cooling},
  author={Feng, Jia-Jin and Wu, Biao and Wilczek, Frank},
  journal={Physical Review A},
  volume={105},
  number={5},
  pages={052601},
  year={2022},
  publisher={APS}
}

@article{koppinen2009phonon,
  title={Phonon cooling of nanomechanical beams with tunnel junctions},
  author={Koppinen, PJ and Maasilta, IJ},
  journal={Physical Review Letters},
  volume={102},
  number={16},
  pages={165502},
  year={2009},
  publisher={APS}
}

@article{wei2008ultrasens,
  title={Ultrasensitive hot-electron nanobolometers for terahertz astrophysics},
  author={Wei, Jian and Olaya, David and Karasik, Boris S and Pereverzev, Sergey V and Sergeev, Andrei V and Gershenson, Michael E},
  journal={Nature Nanotechnology},
  volume={3},
  number={8},
  pages={496--500},
  year={2008},
  publisher={Nature Publishing Group UK London}
}

@article{clements2026sub,
  title={Sub-Doppler cooling of a trapped ion in a phase-stable polarization gradient},
  author={Clements, Ethan and Knollmann, Felix W and Corsetti, Sabrina and Li, Zhaoyi and Hattori, Ashton and Notaros, Milica and Swint, Reuel and Sneh, Tal and Kim, May E and Leu, Aaron D and others},
  journal={Physical Review Letters},
  volume={136},
  number={2},
  pages={023201},
  year={2026},
  publisher={APS}
}

@article{teufel2011sideband,
  title={Sideband cooling of micromechanical motion to the quantum ground state},
  author={Teufel, John D and Donner, Tobias and Li, Dale and Harlow, Jennifer W and Allman, MS and Cicak, Katarina and Sirois, Adam J and Whittaker, Jed D and Lehnert, Konrad W and Simmonds, Raymond W},
  journal={Nature},
  volume={475},
  number={7356},
  pages={359--363},
  year={2011},
  publisher={Nature Publishing Group UK London}
}

@article{riviere2011optom,
  title={Optomechanical sideband cooling of a micromechanical oscillator close to the quantum ground state},
  author={Riviere, Remi and Deleglise, Samuel and Weis, Stefan and Gavartin, Emanuel and Arcizet, Olivier and Schliesser, Albert and Kippenberg, Tobias J},
  journal={Physical Review A},
  volume={83},
  number={6},
  pages={063835},
  year={2011},
  publisher={APS}
}

@article{jaehne2008ground,
  title={Ground-state cooling of a nanomechanical resonator via a Cooper-pair box qubit},
  author={Jaehne, Konstanze and Hammerer, Klemens and Wallquist, Margareta},
  journal={New Journal of Physics},
  volume={10},
  number={9},
  pages={095019},
  year={2008},
  publisher={IOP Publishing}
}

@article{Birnbaum2025photon,
  title={Photon blockade in an optical cavity with one trapped atom},
  author={Birnbaum, K. M. and Boca, A. and Miller, R. and Boozer, A. D. and Northup, T. E. and Kimble, H. J.},
  journal={Nature},
  volume={436},
  number={7047},
  pages={87--90},
  year={2025},
  publisher={Nature Publishing Group UK London}
}

@article{diamandi2025optom,
  title={Optomechanical control of long-lived bulk acoustic phonons in the quantum regime},
  author={Diamandi, Hilel Hagai and Luo, Yizhi and Mason, David and Kanmaz, Tevfik Bulent and Ghosh, Sayan and Pavlovich, Margaret and Yoon, Taekwan and Behunin, Ryan and Puri, Shruti and Harris, Jack GE and others},
  journal={Nature Physics},
  volume={21},
  number={9},
  pages={1482--1488},
  year={2025},
  publisher={Nature Publishing Group UK London}
}

@article{marshall2003towards,
  title={Towards quantum superpositions of a mirror},
  author={Marshall, William and Simon, Christoph and Penrose, Roger and Bouwmeester, Dik},
  journal={Physical Review Letters},
  volume={91},
  number={13},
  pages={130401},
  year={2003},
  publisher={APS}
}

@article{ferreri2024phonon,
  title={Phonon-photon conversion as mechanism for cooling and coherence transfer},
  author={Ferreri, Alessandro and Bruschi, David Edward and Wilhelm, Frank K and Nori, Franco and Macr{\`\i}, Vincenzo},
  journal={Physical Review Research},
  volume={6},
  number={2},
  pages={023320},
  year={2024},
  publisher={APS}
}

@article{aspelmeyer2014cavity,
  title={Cavity optomechanics},
  author={Aspelmeyer, Markus and Kippenberg, Tobias J and Marquardt, Florian},
  journal={Reviews of Modern Physics},
  volume={86},
  number={4},
  pages={1391--1452},
  year={2014},
  publisher={APS}
}

@article{braginski1967ponder,
  title={Ponderomotive effects of electromagnetic radiation},
  author={Braginski, VB and Manukin, AB},
  journal={Sov. Phys. JETP},
  volume={25},
  number={4},
  pages={653--655},
  year={1967}
}

@article{dorsel1983optical,
  title={Optical bistability and mirror confinement induced by radiation pressure},
  author={Dorsel, A and McCullen, John D and Meystre, Pierre and Vignes, E and Walther, H},
  journal={Physical Review Letters},
  volume={51},
  number={17},
  pages={1550},
  year={1983},
  publisher={APS}
}

@article{kippenberg2008cavity,
  title={Cavity optomechanics: back-action at the mesoscale},
  author={Kippenberg, Tobias J and Vahala, Kerry J},
  journal={Science},
  volume={321},
  number={5893},
  pages={1172--1176},
  year={2008},
  publisher={American Association for the Advancement of Science}
}

@article{marquardt2007quantum,
  title={Quantum theory of cavity-assisted sideband cooling of mechanical motion},
  author={Marquardt, Florian and Chen, Joe P and Clerk, Aashish A and Girvin, SM},
  journal={Physical Review Letters},
  volume={99},
  number={9},
  pages={093902},
  year={2007},
  publisher={APS}
}

@article{wilson2007theory,
  title={Theory of ground state cooling of a mechanical oscillator using dynamical backaction},
  author={Wilson-Rae, Ignacio and Nooshi, Nima and Zwerger, W and Kippenberg, Tobias J},
  journal={Physical Review Letters},
  volume={99},
  number={9},
  pages={093901},
  year={2007},
  publisher={APS}
}

@article{dobrindt2008parametric,
  title={Parametric normal-mode splitting in cavity optomechanics},
  author={Dobrindt, Jens M and Wilson-Rae, I. and Kippenberg, Tobias J},
  journal={Physical Review Letters},
  volume={101},
  number={26},
  pages={263602},
  year={2008},
  publisher={APS}
}

@article{wilson2008cavity,
  title={Cavity-assisted backaction cooling of mechanical resonators},
  author={Wilson-Rae, I and Nooshi, N and Dobrindt, Jens and Kippenberg, Tobias J and Zwerger, W},
  journal={New Journal of Physics},
  volume={10},
  number={9},
  pages={095007},
  year={2008}
}

@article{chen2016brillouin,
  title={Brillouin cooling in a linear waveguide},
  author={Chen, Yin-Chung and Kim, Seunghwi and Bahl, Gaurav},
  journal={New Journal of Physics},
  volume={18},
  number={11},
  pages={115004},
  year={2016},
  publisher={IOP Publishing}
}

@article{zhu2023dynamic,
  title={Dynamic {Brillouin} cooling for continuous optomechanical systems},
  author={Zhu, Changlong and Stiller, Birgit},
  journal={Materials for Quantum Technology},
  volume={3},
  number={1},
  pages={015003},
  year={2023},
  publisher={IOP Publishing}
}

@article{otterstrom2018om,
  title={Optomechanical cooling in a continuous system},
  author={Otterstrom, Nils T and Behunin, Ryan O and Kittlaus, Eric A and Rakich, Peter T},
  journal={Physical Review X},
  volume={8},
  number={4},
  pages={041034},
  year={2018},
  publisher={APS}
}

@article{johnson2023laser,
  title={Laser Cooling of Traveling-Wave Phonons in an Optical Fiber},
  author={Johnson, Joel N and Haverkamp, Danielle R and Ou, Yi-Hsin and Kieu, Khanh and Otterstrom, Nils T and Rakich, Peter T and Behunin, Ryan O},
  journal={Physical Review Applied},
  volume={20},
  number={3},
  pages={034047},
  year={2023},
  publisher={APS}
}

@article{blazquez2024oa,
  title={Optoacoustic cooling of traveling hypersound waves},
  author={Bl{\'a}zquez Mart{\'\i}nez, Laura and Wiedemann, Philipp and Zhu, Changlong and Geilen, Andreas and Stiller, Birgit},
  journal={Physical Review Letters},
  volume={132},
  number={2},
  pages={023603},
  year={2024},
  publisher={APS}
}

@article{martinez2025cavity,
  title={Cavity-less {Brillouin} strong coupling in a solid-state continuous system},
  author={Mart{\'\i}nez, Laura Bl{\'a}zquez and Zhu, Changlong and Stiller, Birgit},
  journal={arXiv preprint arXiv:2507.08673},
  year={2025}
}

@inbook{breuer2002theory,
  title={The Theory of Open Quantum Systems},
  author={Breuer, Heinz-Peter and Petruccione, Francesco},
  year={2002},
  isbn={9780198520634},
  publisher={Oxford University Press},
  chapter = {3}
}

@article{nathan2020universal,
  title={Universal Lindblad equation for open quantum systems},
  author={Nathan, Frederik and Rudner, Mark S},
  journal={Physical Review B},
  volume={102},
  number={11},
  pages={115109},
  year={2020},
  publisher={APS}
}

@article{agarwal2013multimode,
  title={Multimode phonon cooling via three-wave parametric interactions with optical fields},
  author={Agarwal, G. S. and Jha, Sudhanshu S.},
  journal={Physical Review A},
  volume={88},
  number={1},
  pages={013815},
  year={2013},
  publisher={APS}
}

@article{smith1972optical,
  title={Optical power handling capacity of low loss optical fibers as determined by stimulated {Raman} and {Brillouin} scattering},
  author={Smith, Richard G},
  journal={Applied Optics},
  volume={11},
  number={11},
  pages={2489--2494},
  year={1972},
  publisher={Optical Society of America}
}

@inbook{Agrawal2013,
    author = {Govind P. Agrawal},
    title = {Nonlinear Fiber Optics},
    publisher = {Elsevier Science},
    year = {2013},
    isbn = {978-0-12-397023-7},
    edition = {Fifth},
    chapter = {9.2}
}

@article{wang2023non,
  title={{Non-Hermitian} optics and photonics: from classical to quantum},
  author={Wang, Changqing and Fu, Zhoutian and Mao, Wenbo and Qie, Jinran and Stone, A Douglas and Yang, Lan},
  journal={Advances in Optics and Photonics},
  volume={15},
  number={2},
  pages={442--523},
  year={2023},
  publisher={Optica Publishing Group}
}

@article{miri2019exceptional,
  title={Exceptional points in optics and photonics},
  author={Miri, Mohammad-Ali and Alu, Andrea},
  journal={Science},
  volume={363},
  number={6422},
  pages={eaar7709},
  year={2019},
  publisher={American Association for the Advancement of Science}
}

@article{zhang2019quantum,
  title={Quantum noise theory of exceptional point amplifying sensors},
  author={Zhang, Mengzhen and Sweeney, William and Hsu, Chia Wei and Yang, Lan and Stone, AD and Jiang, Liang},
  journal={Physical Review Letters},
  volume={123},
  number={18},
  pages={180501},
  year={2019},
  publisher={APS}
}

@article{el2018non,
  title={{Non-Hermitian} physics and {PT} symmetry},
  author={El-Ganainy, Ramy and Makris, Konstantinos G and Khajavikhan, Mercedeh and Musslimani, Ziad H and Rotter, Stefan and Christodoulides, Demetrios N},
  journal={Nature Physics},
  volume={14},
  number={1},
  pages={11--19},
  year={2018},
  publisher={Nature Publishing Group UK London}
}

@article{yang2025optical,
  title={Optical frequency-dependent opposite effective acoustic velocity dispersion in stimulated {Brillouin} scattering},
  author={Yang, Juntong and Huang, Yuelang and Wang, Yuan and Chen, Liang and Bao, Xiaoyi},
  journal={APL Photonics},
  volume={10},
  number={5},
  year={2025},
  publisher={AIP Publishing}
}

@article{fischer2025brillouin,
  title={{Brillouin--Mandelstam} scattering-based cooling of traveling acoustic waves from cryogenic temperatures},
  author={Fischer, Lisa and Bl{\'a}zquez Mart{\'\i}nez, Laura and Zhu, Changlong and Chenevi{\`e}re, Robin and Troles, Johann and Stiller, Birgit},
  journal={Optics Letters},
  volume={51},
  number={1},
  pages={121--124},
  year={2025},
  publisher={Optica Publishing Group}
}

@article{zhang2023quantum,
  title={Quantum coherent control in pulsed waveguide optomechanics},
  author={Zhang, Junyin and Zhu, Changlong and Wolff, Christian and Stiller, Birgit},
  journal={Physical Review Research},
  volume={5},
  number={1},
  pages={013010},
  year={2023},
  publisher={APS}
}

@article{van2016unifying,
  title={Unifying {Brillouin} scattering and cavity optomechanics},
  author={Van Laer, Rapha{\"e}l and Baets, Roel and Van Thourhout, Dries},
  journal={Physical Review A},
  volume={93},
  number={5},
  pages={053828},
  year={2016},
  publisher={APS}
}

@article{tomasella2025strong,
  title={Strong optomechanical coupling at room temperature with a centimeter-scale quartz crystal},
  author={Tomasella, Davide and Tarrago Velez, Santiago and Nielsen, Sissel Bay and Van der Heijden, Joost and Hoff, Ulrich Busk and Andersen, Ulrik Lund},
  journal={Physical Review Applied},
  volume={23},
  number={5},
  pages={054024},
  year={2025},
  publisher={APS}
}

\end{document}